\documentclass[aip,apl,amsmath,amssymb,reprint]{revtex4-1}

\usepackage{graphicx}
\usepackage{dcolumn}
\usepackage{bm}
\usepackage[utf8]{inputenc}
\usepackage[T1]{fontenc}
\usepackage{mathptmx}
\usepackage{etoolbox}
\usepackage{braket}
\usepackage{bm}
\usepackage{colortbl}
\makeatletter
\def\@email#1#2{%
 \endgroup
 \patchcmd{\titleblock@produce}
  {\frontmatter@RRAPformat}
  {\frontmatter@RRAPformat{\produce@RRAP{*#1\href{mailto:#2}{#2}}}\frontmatter@RRAPformat}
  {}{}
}%
\makeatother
\begin{document}

\preprint{AIP/123-QED}

\title{Scalar Gravitational Aharonov-Bohm Effect: Generalization of the Gravitational Redshift}
\author{Michael E. Tobar}
 \email{michael.tobar@uwa.edu.au}
 \affiliation{ Quantum Technologies and Dark Matter Labs, Department of Physics, University of Western Australia, Crawley, WA 6009, Australia.}
 \author{Michael T. Hatzon}
 \affiliation{ Quantum Technologies and Dark Matter Labs, Department of Physics, University of Western Australia, Crawley, WA 6009, Australia.} 
 \author{Graeme R. Flower}
 \affiliation{ Quantum Technologies and Dark Matter Labs, Department of Physics, University of Western Australia, Crawley, WA 6009, Australia.}
  \author{Maxim Goryachev}
 \affiliation{ Quantum Technologies and Dark Matter Labs, Department of Physics, University of Western Australia, Crawley, WA 6009, Australia.}

\begin{abstract}
The Aharonov-Bohm effect is a quantum mechanical phenomenon that demonstrates how potentials can have observable effects even when the classical fields associated with those potentials are absent. Initially proposed for electromagnetic interactions, this effect has been experimentally confirmed and extensively studied over the years. More recently, the effect has been observed in the context of gravitational interactions using atom interferometry. Additionally, recent predictions suggest that temporal variations in the phase of an electron wave function will induce modulation sidebands in the energy levels of an atomic clock, solely driven by a time-varying scalar gravitational potential\cite{Chiao2023Gravitational}. In this study, we consider the atomic clock as a two-level system undergoing continuous Rabi oscillations between the electron's ground and excited state. We assume the photons driving the transition are precisely frequency-stabilized to match the transition, enabling accurate clock comparisons. Our analysis takes into account, that when an atom transitions from its ground state to an excited state, it absorbs energy, increasing its mass according to the mass-energy equivalence principle. Due to the mass difference between the two energy levels, we predict that an atomic clock in an eccentric orbit will exhibit a constant frequency shift relative to a ground clock corresponding to the orbit's average gravitational redshift, with additional modulation sidebands due to the time-varying gravitational potential.
\end{abstract}

\maketitle

The original proposal for the electromagnetic Aharonov-Bohm (AB) effect \cite{AB1959} focused on the scalar and vector potentials of the electromagnetic interaction. The magnetic effect is a phenomenon where a charged particle's wave function is affected by the magnetic vector potential, $\vec{A}$, even when both the electric and magnetic fields are zero \cite{AB1959}. Underlying this effect is the general concept of geometric phase \cite{Berry1984}, which is apparent in many areas of physics \cite{Wilczek1989}, including condensed matter physics \cite{Resta_2000,Xiao2007}, optics \cite{Chiao1990,Lipson:90}, fluid mechanics \cite{Perrot:2019bb}, etcetera. On the other hand, the original work also proposed the scalar-electric AB effect, where the electric scalar potential, $V$, creates a geometric phase over time, with zero spatial variation so the electric field is zero. Since the experimental set-up for the vector-magnetic AB effect is much easier to realize, there have been several tests of the vector-magnetic AB effect \cite{chambers,tono}. However, no clean scalar-electric AB effect has ever been implemented \cite{electric-ab}. 

In the standard scalar-electric AB proposal, charges are sent along different paths with a potential difference between them but with no electric fields\cite{AB1959}. The observational signature is a shift in the quantum interference pattern of charges, which evolves with time as the two path lengths undergo different potentials \cite{AB1959}. In contrast, a more recent proposal \cite{Chiao23} places a quantum system inside a Faraday cage with a time-varying scalar potential, so one effectively compares the quantum system as the potential is turned on and off (a chopping experiment), which highlighs the temporal nature of the scalar effect. In an analagous way to the AC Stark effect \cite{Delone:1999aa}, this setup calculates that several energy side-bands in the spectrum of the quantum system will be produced, rather than a shift of interference fringes as in the originally proposed set-up \cite{AB1959}.

Since the gravitational potential is a Newtonian scalar potential, its analog in the gravitational sector closely resembles the scalar electromagnetic AB effect as discussed in \cite{Chiao23,Chiao2023Gravitational}. A recent experimental verification of the scalar gravitational AB effect \cite{overstreet,Hohensee2012} employed a method similar to the original AB experiment \cite{AB1959}. In this setup, a matter beam was split into two paths, with one path experiencing a different gravitational potential to the other, resulting in a shift in the interference pattern when the beams were recombined. 

The analog temporal setup to measure the scalar gravitational AB effect was recently proposed in \cite{Chiao2023Gravitational}, which considered a quantum system in orbit around a massive body. To be sensitive to the gravitational AB effect, the orbit must have non-zero eccentricity, so geometric phase effects related to the time-varying gravitation potential will be present. Since the quantum system will be in free-fall, by the equivalence principle the gravitational field is locally screened.  However, the time-varying gravitational potential will change the energy levels of a quantum system, which should develop sidebands, which are harmonics of the orbit frequency, which are the signature of the scalar Aharonov-Bohm effect. 

\begin{figure}[t]
 \includegraphics[width=0.43\textwidth]{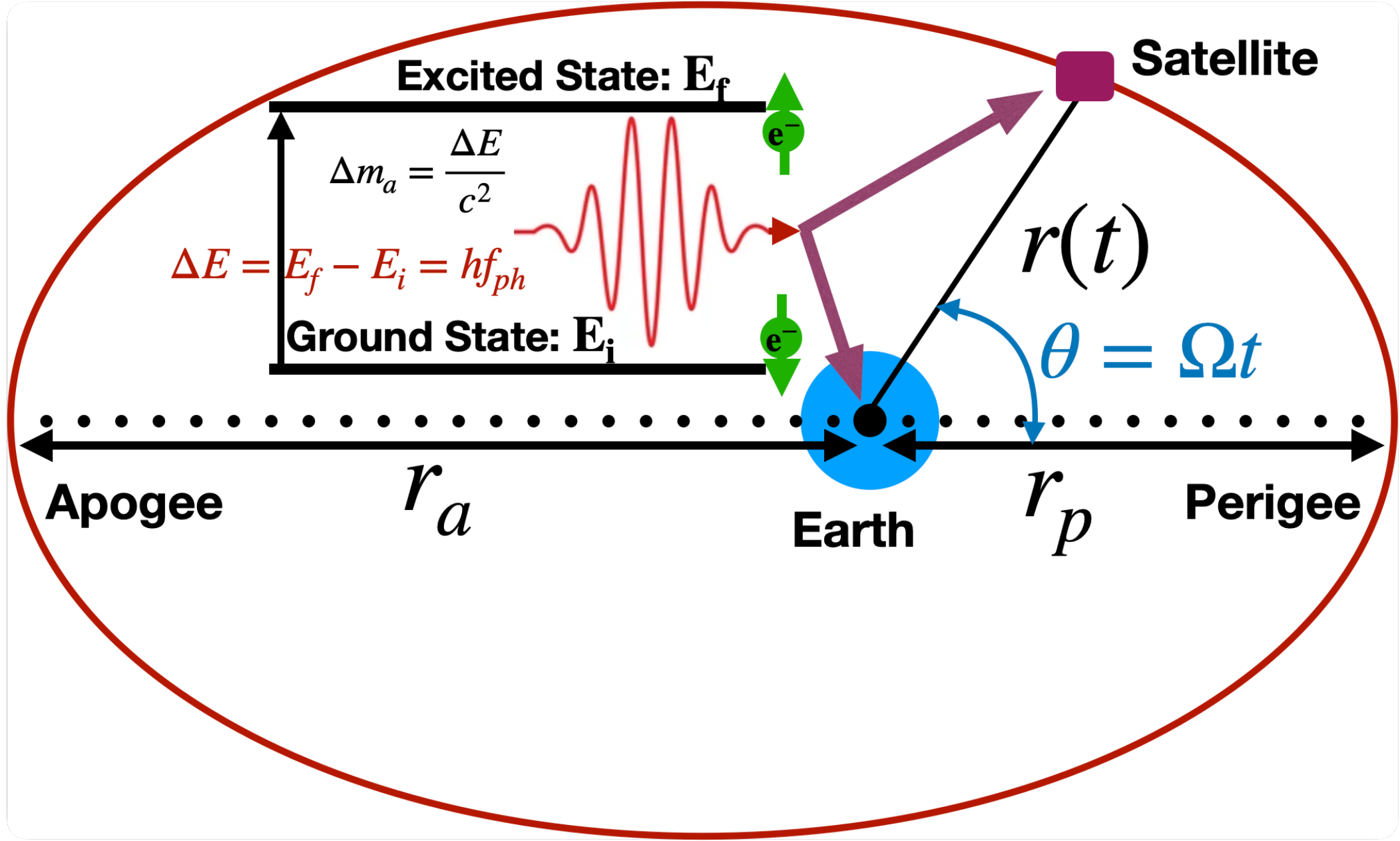}
	 \caption{Keplerian orbit of a satellite with an onboard atomic clock, showing the temporal variation of the satellite position from the center of the Earth, $r(t)$, perigee ($r_p$) occurs at $t=0$ and apogee ($r_a$) at $t=\pi/\Omega$. Inset: Energy level diagram of an atomic clock based on a hyperfine transition of energy $\Delta E=E_f-E_i$, where the ground state the electron spin, of energy $E_i$, is anti-aligned with the spin of the nucleus while the excited state of energy $E_f$ is aligned. The transition from ground state to excited state absorbs a photon of energy, $hf_{ph}=\Delta E$, and the atom will increase its mass by $\Delta m_a=\Delta E/c^2$.}
     \label{KOrbit}
   \end{figure}

Experiments that have the potential to measure such geometric phase effects include precision atomic clocks in space, such as the Atomic Clock Ensemble in Space (ACES) mission \cite{aces} on board the International Space Station (ISS), as well as other missions that propose optical clocks in space \cite{Gou21,Derevianko:2022aa,Schkolnik:2023aa}. Also, data from Galileo clocks in orbits with non-zero eccentricity \cite{Herrmann2018,Delva2018} could also exhibit a signal. This means that instead of measuring an interference pattern in an interferometric setup, the focus of this type of experiment is to measure the energy shifts in the quantum system caused by the time-varying gravitational potential. In this type of experiment, comparisons are made with atomic clock transition onboard a satellite with one on Earth, as illustrated in Fig. \ref{KOrbit}. The ground clocks experience a relatively constant gravitational potential, whereas the orbiting clocks are subject to a time-varying potential. Currently, most atomic clocks in space are microwave-based, typically using hydrogen, rubidium, or cesium atoms. These clocks rely on hyperfine transitions, where the ground state has an electron anti-aligned with the nucleus and the excited state has it aligned, as depicted in Fig. \ref{KOrbit}. 

In practice, when an oscillating electromagnetic field has a frequency, $f_{ph}$, close to the transition frequency of an atomic clock, all other energy levels can be ignored. Thus, we approximate an atomic clock as a two-level system with a ground state, $\ket{\psi_{i}}$, an excited state $\ket{\psi_{f}}$, and a photon field precisely tuned to the transition frequency, derived from the energy difference between these two levels, by $\Delta E=E_f-E_i=hf_{ph}$. The resulting dynamics are well known, characterized by oscillations between the ground and excited states at the Rabi frequency \cite{Rabi1938,Ramsey62}. In this section, we derive the effect of an additional time-dependent gravitational AB phase on a generic two-level system. In general, it has been shown that the gravitational AB phase, $\varphi _g (t)$, for a transition involving the excitation of an atomic bound electron is given by \cite{Chiao2023Gravitational}, 
\begin{equation}
    \varphi _g (t) = \frac{m_e}{\hbar} \int _0 ^t \Phi _g (t') dt' ~,
    \label{AB-grav}
\end{equation}
where, $m_e$ is the mass of the electron, and the gravitational potential, $\Phi _g (t)$, is in general time varying \cite{Chiao2023Gravitational}. The AB phase introduces an additional phase term that alters the standard dynamics of Rabi oscillations. In our approach, we incorporate the AB phase by assuming the quasi-static limit, where the variations occur at a much lower frequency compared to the Rabi oscillations. This means the changes are slow and adiabatic, allowing the system to remain in its instantaneous eigenstate throughout the process. 

The effect of the time varying gravitational AB phase on the energy levels of a two level clock system (ignoring kinetic Doppler terms) may be written as, 
\begin{equation}
E_{i}(t)=E_{i0}+m_e\Phi(t)~~\text{and} ~~ E_{f}(t)=E_{f0}+m_e^{*}\Phi(t),
\end{equation} 
where $E_{i0}$ and $E_{f0}$ are respectively the ground and excited state energy of the two-level system in the absence of a gravitational potential. When an atom transitions from its ground state to an excited state, it absorbs energy.  According to the mass-energy equivalence principle, there is a change in the atom's mass through $\Delta E=\Delta m_a c^2$, this additional energy contributes to the mass of the atom. Thus, in this case, our excited electron wave function effectively gains a mass of $m_e^*=m_e+\Delta m_a$ due to its increased energy in its excited state. The increase in mass is equivalent to the absorbed photon's kinetic mass, $m_{ph}$, so $\Delta m_a=m_{ph}=hf_{ph}/c^2$. 

We can then formulate a Hamiltonian of the form 
\begin{equation}
H(t)=(E_{i0}+m_e\Phi(t))\ket{\psi_{i}}\bra{\psi_{i}}+(E_{f0}+m_e^{*}\Phi(t))\ket{\psi_{f}}\bra{\psi_{f}}.
\end{equation}
To evaluate the time-dependant wave function, we begin by calculating the propagator 
\begin{equation}
U(t)=e^{-\frac{i}{\hbar}\int_{0}^{t}(E_{i0}+m_e\Phi(t'))\ket{\psi_{i}}\bra{\psi_{i}}+(E_{f0}+m_e^{*}\Phi(t'))\ket{\psi_{f}}\bra{\psi_{f}}dt'}.
\end{equation}
Identifying the gravitational AB phase terms given by \eqref{AB-grav}, we may then write the propagator as,
\begin{equation}
U(t)=e^{-i(\frac{E_{i0}t}{\hbar}+\varphi_{g}(t))\ket{\psi_{i}}\bra{\psi_{i}}-i(\frac{E_{f0}t}{\hbar}+\varphi_{g}^*(t))\ket{\psi_{f}}\bra{\psi_{f}}},
\end{equation}
where $\varphi_g^*(t)$ is similar to \eqref{AB-grav}, but with $m_e\rightarrow m_e^*$. Then writing the propagator exponential as power series, we find that,
\begin{multline}
U(t)= \\*
\Sigma_{k=0}^{\infty}\frac{(i(-\frac{E_{i}t}{\hbar}-\varphi_{g}(t))\ket{\psi_{i}}\bra{\psi_{i}}+i(-\frac{E_{f}t}{\hbar}-\varphi_{g}^*(t))\ket{\psi_{f}}\bra{\psi_{f}})^{k}}{k!}.
\end{multline}
Because the eigenfunctions are orthogonal, the cross terms will vanish and we obtain,
\begin{multline}
U(t)=\Sigma_{k=0}^{\infty} \frac{(i(-\frac{E_{i0}t}{\hbar}-\varphi_{g}(t)))^{k}}{k!}\ket{\psi_{i}}\bra{\psi_{i}}\\
+\frac{(i(-\frac{E_{f0}t}{\hbar}-\varphi_{g}^*(t)))^{k}}{k!}\ket{\psi_{f}}\bra{\psi_{f}},
\end{multline}
then collecting the power series as individual exponentials we obtain,
\begin{equation}
U(t)=e^{i(-\frac{E_{i0}t}{\hbar}-\varphi_{g}(t))}\ket{\psi_{i}}\bra{\psi_{i}}+e^{i(-\frac{E_{f0}t}{\hbar}-\varphi_{g}^*(t))}\ket{\psi_{f}}\bra{\psi_{f}}.
\end{equation}
So, the final wave function is \begin{equation}
\psi(t)=U(t)\psi(0),
\end{equation} where we take $\psi(0)=c_{i}\ket{\psi_{i}}+c_{f}\ket{\psi_{f}}$, the initial wave function, a general superposition of ground and excited states, which defines the state of the Rabi oscillation at t=0.
Hence, 
\begin{multline}
\psi(t)=e^{i(-\frac{E_{i}t}{\hbar}-\varphi_{g}(t))}c_{i}\ket{\psi_{i}}\bra{\psi_{i}}\ket{\psi_{i}}\\
+e^{i(-\frac{E_{f}t}{\hbar}-\varphi_{g}^*(t))}c_{f}\ket{\psi_{f}}\bra{\psi_{f}}\ket{\psi_{f}},
\end{multline}
which becomes,
\begin{equation}
\psi(t)=e^{i(-\frac{E_{i0}t}{\hbar}-\varphi_{i}(t))}c_{i}\ket{\psi_{i}}+e^{i(-\frac{E_{f0}t}{\hbar}-\varphi_{g}^*(t))}c_{f}\ket{\psi_{f}}.
\end{equation}
Taking out a common mode phase term, we may write,
\begin{equation}
\psi(t)=e^{i(-\frac{E_{i0}t}{\hbar}-\varphi_{i}(t))}(c_{i}\ket{\psi_{i}}+e^{i(-\frac{\Delta E_{0}}{\hbar}t-(\varphi_{g}^*(t)-\varphi_{g}(t)))}c_{f}\ket{\psi_{f}}).
\end{equation}
where $\Delta E_{0}=E_{f0}-E_{i0}=hf_{ph0}$, is the energy difference between the two states under zero gravitational potential, and $f_{ph0}$ is the clock frequency under the same condition. Then by taking the rotating wave approximation, and using the fact that a global phase is known to not be observable, in the rotating frame we may write,
\begin{equation}
\psi(t)=(c_{i}\ket{\psi_{i}}+e^{-i(\frac{\Delta E_{0}t}{\hbar}+\Delta\varphi_{g}(t))}c_{f}\ket{\psi_{f}}),
\end{equation}
where $\Delta\varphi_{g}(t)=\varphi_{g}^*(t)-\varphi_{g}(t)$ may be regarded as a time dependent gravitational mixing angle between the two states. Thus, we may regard the ground state as time independent, with the excited state undergoing a temporal AB phase shift due to the time varying gravitational potential.

Assuming our atomic clock is in a satellite in an elliptical orbit so that the distance between the satellite and center of the Earth changes with time, the gravitational potential will be of the form, $\Phi _g (t)=-G M~r(t)^{-1}$, where $G$ is Newton's constant, $M$ is the mass of the Earth about which our quantum system will orbit, and $r (t)$ is the time-dependent distance between the satellite and one focus of the orbit {\it i.e.} the center of the Earth. Since our quantum system is in orbit (free fall) in a satellite, then by the equivalence principle the quantum system will be locally in a zero gravitational field, {\it i.e.} the field has been transformed away by going to a free falling frame. Assuming the satellite's trajectory as a Keplerian orbit (as shown in Fig.\ref{KOrbit}), then $r(t)^{-1}=r_0^{-1}+\left(A / r_0^2\right) \cos(\Omega t)$, where $r_0=(r_a+r_p)/2$ is the average value of the apogee, $r_a$ and perigee, $r_p$, and the value of $A=(r_a-r_p)/2$ is the average difference.

Thus for a Keplerian orbit, the temporal variation of the gravitational potential will be off the form.
\begin{equation}
    \label{AB-pot}
	\Phi_g(t)=-\frac{G M}{r_0}\left(1+e\cos(\Omega t)\right),
\end{equation}
where $e=\frac{A}{r_0}$, which is the same as derived in \cite{Chiao2023Gravitational} but without any approximations.

In this case the time varying gravitational mixing angle becomes,
\begin{equation}
\begin{aligned}
\Delta \varphi _{g}(t) =- \frac{\Delta E_0GM}{\hbar r_0c^2} \int _0 ^t \left(1+e\cos(\Omega t')\right) dt'=\alpha\sin(\Omega t).
\end{aligned}
    \label{AB-grav2}
\end{equation}
where  
\begin{equation}
\alpha \equiv e\left(\frac{GM}{r_0c^2}\right)\left(\frac{2\pi f_{ph0}}{\Omega}\right), 
\label{alpha}
\end{equation}
is the dimensionless gravitational depth of modulation. 
Following the same process as in \cite{Chiao2023Gravitational}, which solves the time dependent Schr{\"o}dinger equation, the solution to the wave function becomes,
\begin{equation}
\begin{aligned}
& \Psi_f(\mathbf{r}, t) \\
& =\Psi_f(\mathbf{r}) \sum_{n=-\infty}^{\infty}(-1)^n J_n(\alpha) \exp \left(-\frac{i\left(\Delta E_0(1+\frac{G M}{r_0c^2})-n \hbar \Omega\right) t}{\hbar}\right)
\label{wf}
\end{aligned}
\end{equation}
From (\ref{wf}) we can see that each energy level between the ground and excited state splits into a multiplet $\Delta E^{(n)}$ given by,
\begin{equation}
\Delta E^{(n)}=\Delta E_0(1+ \frac{GM}{r_0c^2})\pm n\hbar \Omega  \equiv \Delta E \pm n\hbar \Omega ,~~\text{ with}\ n, \text{an integer}.
\label{energy-2-g}
\end{equation}
The result in \eqref{energy-2-g} has a constant shift of $\Delta E=\Delta E_0(1+\frac{GM}{r_0c^2})$, which is equivalent to the energy shift between the two atomic levels due to the average gravitational potential of the orbit. Note that the gravitation potential actually causes a blue shift with respect to the atomic transition under no gravitational potential (i.e. at r = $\infty$), but is red shifted with respect to the atomic clock on Earth.

Note, if $\alpha=0$ only the $n=0$ part of the expansion in \eqref{wf} survives and there are no sidebands, and if  $\alpha<<1$ only the $n=\pm1$ sidebands are significant, and equivalent to a single tone frequency modulation, which will occur for extremely small eccentricities. When $\alpha>1$ over-modulation occurs and extra sidebands are created, which may be searched for as a sign of this effect. In the case when $\alpha>>1$, one finds suppressed sidebands for $n << \alpha$, until at $n\sim n_{max }\approx \alpha$, there is a sharp up shoot in the value of $J_n(\alpha)$, and hence the power in these sidebands. In this case for $n> n_{max}$, $J_n(\alpha)$ exponentially decays to zero, so that sidebands beyond $n_{max}$ do not contribute \cite{Chiao2023Gravitational}. Thus, \eqref{energy-2-g} may be thought of as a generalization of the gravitational redshift, and predicts extra sidebands depending on the modulation parameter, $\alpha$, given in \eqref{alpha}.

The  Atomic Clock Ensemble in Space (ACES) mission \cite{aces} will place microwave atomic clocks on the International Space Station (ISS), which could measure this effect. The International Space Station is an almost circular, low Earth orbit with orbital parameters given by: (i) perigee and apogee radius from the center of the Earth are $r_p = 6.800 \times 10^6$ m and $r_a = 6.810 \times 10^6$ m, respectively, which corresponds to a perigee altitude of 400 km and apogee altitude of 410 km given that the Earth's radius is $r_E \approx 6400$ km. These values give an eccentricity of $e_{ISS}=7.34\times10^{-4}$. The period of a satellite with this apogee/perigee is about $T_{ISS} \approx$ 90 minutes or 5400 seconds giving an angular frequency of $\Omega_{ISS} = \frac{2 \pi}{T_{ISS}}  = 1.2 \times 10^{-3} \rm{\frac{rad}{sec}}$ (or frequency of $f_{ISS}=1.85 \times 10^{-4} ~ {\rm Hz}$). Substituting these values into \eqref{alpha} gives $\alpha_{ISS}=2.6\times10^{-9}f_{ph}$, where $f_{ph}$ is in Hz. Two types of atomic clocks onboard the ISS, an active hydrogen maser of frequency 1.42 GHz and the PHARAO (Projet d'Horloge Atomique par Refroidissement d'Atomes en Orbite) cesium clock of frequency 9.19263177 GHz \cite{Laurent:2020aa}, which will have modulation sidebands created through the gravitational modulation index of $\alpha_H=3.7$ and $\alpha_{Cs}=23.8$ respectively.

Data from Global Positioning Systems clocks have already been used to test fundamental physics \cite{Wolf97,Derevianko}, such as tests on special relativity and the hunt for dark matter. The Galileo satellites, part of the European Space Agency's Galileo navigation system, mostly follow nearly perfect circular orbits \cite{Kouba:2019aa}. However, two satellites were accidentally placed in elliptical orbits, and provided an opportunity to achieve the best limits on measuring gravitational redshift by comparing the onboard hydrogen maser clocks with ground-based clocks \cite{Herrmann2018,Delva2018}. It is also possible that the effects of the gravitational AB effect exists in this data, as the DC term in \eqref{energy-2-g} is equivalent to the redshift, which was measured in \cite{Herrmann2018,Delva2018} at a precision of parts in $10^5$, while the Aharonov-Bohm phase predicts extra sidebands. The orbital period of the satellite is 12.94 hours (21.5 $\mu$Hz) with an eccentricity of $e_G=0.162$, with a perigee and apogee radius from the center of the Earth of $r_p =  23,445 km$ and $r_a = 32,510 km$, respectively, so $r_0=27,977.5$ km, and $A_G=453.2$ km and thus the gravitational modulation parameter for the Galileo orbit is $\alpha_G =1.2\times10^{-6}f_{ph}$ where $f_{ph}$ is in Hz. For the hydrogen maser clock transition at 1.42 GHz, $\alpha_g\sim 1699$, with a predicted peak modulation sideband at around 36.5 mHz. Note, that this orbit causes a modulation, $\alpha_G$, which is about 460 times greater than the modulation index induced by the ISS orbit. To be able to measure this peak frequency sideband and avoid aliasing, by the Nyquist–Shannon sampling theorem, a comparison measurement must be made at twice the rate of frequency (i.e. at 73 mHz) or less. This means the averaging time set for the comparison for a single measurement should be on the order of 14 seconds or less.

We have extended the original proposal for an experimental tests of the gravitational AB effect \cite{Chiao2023Gravitational}, which predicted that a time-varying gravitational potential will cause energy level shifts of an atomic clock transition. We have shown that the predicted phenomena can be viewed as a generalization of the gravitational redshift, which predicts the correct constant value of the redshift, plus additional sidebands described by the Jacobi-Anger expansion as originally predicted in \cite{Chiao2023Gravitational}. The experimental verification of the redshift has been a focused test of general relativity over the years \cite{pound,Vessot1980,Wolf97,Derevianko}, in such experiments, additional Doppler effects due to the velocity of the satellite are known to reduce the measurable signal but do not cancel it. Therefore, Doppler shifts must be properly accounted for to accurately determine the measurable gravitational effect  for clock comparisons between one in an eccentric orbit and a stationary ground clock, and remains an area for future research.

\subsection*{Acknowledgements}
The Author's would like to thank Andrei Derevianko, Doug Singleton, Nathan Inan and Raymond Chiao for valuable discussions on this topic. This work was funded by the Australian Research Council Centre of Excellence for Engineered Quantum Systems, CE170100009 and  the Australian Research Council Centre of Excellence for Dark Matter Particle Physics, CE200100008.

\end{document}